\documentclass[useAMS,usenatbib,usegraphicx]{mn2e}
\usepackage{times}

\title[Radio structure of the most distant radio-detected quasar
at the ten milli-arcsecond scale]
{Radio structure of the most distant radio-detected quasar
at the ten milli-arcsecond scale}
\author[S.~Frey et al.]{S.~Frey$^{1}$\thanks{E-mail: frey@sgo.fomi.hu (SF),
mosoni@sgo.fomi.hu (LM), paragi@jive.nl (ZP), lgurvits@jive.nl (LIG)}, L.~
Mosoni$^{1}$, Z.~Paragi$^{2,1}$ and L.I.~Gurvits$^{2}$\\
$^{1}$F\"OMI Satellite Geodetic Observatory, P.O. Box 585, H-1592 Budapest,
Hungary\\
$^{2}$Joint Institute for VLBI in Europe, Postbus 2, 7990 AA Dwingeloo, The
Netherlands
}
\begin{document}

\date{Accepted date. Received date; in original form date}

\pagerange{\pageref{firstpage}--\pageref{lastpage}} \pubyear{2003}

\maketitle

\label{firstpage}

\begin{abstract}
We present a high resolution radio image of SDSS 0836+0054 identified
recently as the most distant radio-detected quasar at a redshift of $z=5.82$.
The observation was carried out with ten antennas of the European VLBI Network,
spread from Europe to China and South Africa, at 1.6~GHz frequency on 2002 June
8. 
The source is detected with a total flux
density of 1.1~mJy, equal to its flux density measured in the VLA FIRST survey.
We found no indication of multiple images produced by gravitational lensing.
The radio structure of the quasar at $\sim10$-mas angular
resolution appears somewhat resolved. It resembles the radio structure typical
for
lower redshift radio-loud active galactic nuclei. We obtained so far
the best astrometric position of the source with an accuracy better than 8~mas,
limited mainly by the structural effects in the phase-reference calibrator
source.
\end{abstract}


\section{Introduction}

The object SDSSp J083643.85+005453.3 (hereafter: SDSS 0836+0054) is the highest
redshift radio-detected quasar known
to date ($z=5.82$) discovered using multicolor imaging data from the Sloan
Digital Sky Survey (SDSS) \citep{fan01}. The object is one of the seven $z>5.7$
quasars known. Six of them form a complete
colour-selected flux-limited sample from an area of 2870~deg$^2$ \citep{fan03}.
The quasar SDSS 0836+0054 was identified with an unresolved radio source within
$1\arcsec$ in the VLA FIRST survey
\citep{whi97}\footnote{http://sundog.stsci.edu}, where
the source total flux density was $1.11\pm0.15$~mJy at 1.4~GHz (21~cm
wavelength). The total flux density in the NRAO VLA Sky 
Survey \citep[NVSS, ][]{con98}\footnote{http://www.cv.nrao.edu/nvss} was
$2.5\pm0.5$~mJy at the same frequency. 
                                                                            
According to the recent studies \citep[e.g. ][ and references therein]{bar02},
$z\sim6$ is close to the reionization epoch of the Universe. In the 
optical spectrum of SDSS 0836+0054, however, the Gunn--Peterson effect is not
seen. There is detectable emission blueward of the Ly$\alpha$ line, i.e. the
intergalactic medium is already highly ionized at $z\sim5.8$ along this line of
sight \citep{fan01}.

The estimated mass of the central black hole in SDSS 0836+0054 is
$4.8\times10^{9} M_{\sun}$, assuming that the quasar is emitting at the
Eddington luminosity and its apparent flux is not significantly magnified by 
beaming or gravitational lensing \citep{fan01}. 
There are, however, indications that the
latter assumptions may not be correct. Simulations show that up to one third of 
$z\sim6$ quasars may be magnified due to gravitational lensing by a factor of 
ten or more. Therefore their black hole masses are overestimated by the same
factor \citep{wyi02a,wyi02b}. The probability estimations strongly depend on the 
quasar luminosity function used. Considering a wider range of possible 
luminosity functions, \citet{com02} conclude that present
data allow lensing probability of essentially 100\%. 
In turn, the luminosity function can be 
constrained using the detections or absence of lensing events with multiple
images of $z\sim6$ quasars.

Black hole mass values of $z\goa6$ quasars have great importance for
hierarchical structure formation models in the early Universe, placing
important constraints on parameters such as the seed mass, velocity dispersion,
radiative efficiency and accretion luminosity. 
The time needed for a supermassive $\sim10^{9} M_{\sun}$ black hole to grow from
a
stellar-mass seed may be comparable to the age of the Universe at this redshift
\citep{hai01}.

The first four $z>5.7$ SDSS quasars \citep{fan01} including SDSS 0836+0054 are
detected in
X-rays \citep{bra02}. Short exploratory Chandra observations revealed that SDSS
0836+0054 is consistent with a point-like X-ray source
\citep{bra02,mat02,sch02,bec03}. Its position agrees with the optical position
within
the astrometric precision of Chandra ($\sim1\arcsec$).
Based on the limited data available, the
broad-band X-ray and optical properties of the extremely distant quasars are
apparently similar to those of lower redshift ones, implying no strong
evolution out to $z\sim6$ \citep{mat02}. On the contrary, \citet{bec03} found
that 
high-redshift quasars are significantly more X-ray quiet than their low-redshift 
counterparts.

Any indication of a milli-arcsecond scale ``core--jet'' radio structure revealed
by Very Long Baseline Interferometry (VLBI) could be considered as a strong
support for the existence of a supermassive black hole powering SDSS 0836+0054.
These structures are common in radio-loud active galactic nuclei (AGN) in the
broadest range of redshifts from $z\sim0.01$ to $z\sim4$ \citep[e.g.
][]{par99,fom00}.
According to the AGN paradigm, the radio emission originates from incoherent
synchrotron emission in compact (parsec-scale) jets in the close vicinity of
the central black hole.  

Here we describe a 1.6-GHz European VLBI Network (EVN) observation of SDSS
0836+0054 and present the high resolution radio image of the quasar. Based on
our phase-reference VLBI observation, we derive the accurate astrometric
position of the object. We conclude that the image of the quasar is unlikely to
be
split by gravitational lensing. Throughout this letter, we adopt a flat
cosmological model with $\Omega_{m}=0.3$, $\Omega_{\Lambda}=0.7$ and
$H_{\rm{0}}=65$~km~s$^{-1}$~Mpc$^{-1}$. In this model, 1 milli-arcsecond (mas)
angular separation corresponds to a linear separation of 6.25~pc at the distance
of SDSS 0836+0054.

\section{Observation and data reduction}

We used ten antennas of the EVN to observe SDSS 0836+0054 at 1.6~GHz (18~cm
wavelength) in phase-reference mode. The radio telescopes and their parameters
are listed in Table~\ref{ant}. Due to the redshift, the
observing frequency corresponds to 11~GHz in the rest frame of the source.

Phase-referencing is a technique of extending the signal coherence time in order
to
increase the sensitivity of the VLBI array. This is done by regularly
interleaving observations between the weak target source and a nearby strong
reference source. Delay, delay rate and phase solutions obtained for the
phase-reference calibrator are interpolated and applied for the target source,
thus removing most of the phase errors introduced by geometric, ionospheric,
atmospheric and instrumental effects \citep[e.g.][]{bea95}. The ability of the
EVN to detect weak (sub-mJy), compact radio sources has been demonstrated
recently \citep[e.g.][]{garrett01,garrington01}.

The observation of SDSS 0836+0054 took place on 2002 June 8. We selected PKS
0837+012 \citep[J0839+0104; $z=1.123$, ][]{owe95} as the phase-reference
calibrator. The angular separation between the target and the reference source 
is $\sim0.8\degr$. The ICRF
coordinates of PKS 0837+012 are available from the VLBA Calibrator
Survey\footnote{http://magnolia.nrao.edu/vlba\_calib} \citep{bea02}, with an
accuracy of 0.35 and 0.58~mas in right ascension and declination, respectively.
In principle it allows us to make a precise determination of the poorly known
position
of SDSS 0836+0054 through differential astrometry inherently provided by the
phase-referencing technique. The source J0825+0309 was used as a fringe-finder.
 
At the frequency of 1.6~GHz, variations in the ionospheric electron
column density dominate the propagation effects affecting the
quality of phase transfer between the target and calibrator
sources. These variations have static and dynamic components, both
spatial and temporal, which limit the maximum achievable
dynamic range ($D$) of VLBI phase-referencing. According to \citet{bea95},
the upper limit of $D$ is given by 
$D \simeq 30 \nu \theta ^{-1} (\Delta I)^{-1}$,
where $\Delta I$ is peak daily residual total electron content (TEC, in
units of $10^{17}$~m$^{-2}$), $\nu$ is the observing frequency in
GHz, $\theta$ is the angular distance between the target and
reference sources (in degrees). Our observation took place
relatively close after the peak of the solar cycle.
Considering the worst case of the maximum TEC as $\Delta I =
5\times 10^{-17}$~m$^{-2}$ \citep[which corresponds to the solar maximum
daytime value, ][]{bea95}, the conservative
estimate of the maximum achievable dynamic range is $D\simeq 12$.
The dynamic component is caused essentially by medium-scale
travelling ionosphere disturbances (MSTIDs). These are
time-variable and highly unpredictable. MSTIDs are known to vary on
the time scale of tens of minutes to hours and linear scale of
hundreds of kilometres \citep[e.g.][]{van92}. At the maximum, which
normally lasts for up to 5\% in time, the amplitude of MSTIDs could
worsen the upper limit of the achievable dynamic range by a factor
of $\sim 2$ at our target--reference separation comparing to the worst
case of static component \citep[e.g.][]{con93}. 
By splitting up the data set into four equal time
intervals, we did not notice any significant difference between our results on
this time scale.                            
                                                                       
The total observing time was 5 hours with dual circular polarization at
all antennas but Sheshan where only left circular polarization was observed.
The data were recorded with the Mark IV VLBI data acquisition system at
the rate of 256 Mbit/s per station with 2-bit sampling, resulting in 64~MHz
total bandwidth. The correlation took place at the 
EVN Data Processor at the Joint Institute for VLBI in Europe (JIVE), Dwingeloo,
The Netherlands.                     

We used 5-min switching cycles between SDSS 0836+0054 and the phase-reference
calibrator quasar J0839+0104. The target source was observed for
$\sim210$-s intervals in each cycle. The total on-source time for SDSS 0836+0054
was 2.9 hours. The achievable 1-$\sigma$ image noise was estimated to be
27~$\mu$Jy/beam.

\begin{table}
     \centering
     \begin{minipage}{90mm}
     \caption{EVN telescopes and their characteristics at 1.6~GHz}
     \label{ant}
     \begin{tabular}{@{}lcc}
     \hline
     Radio telescope (Country) & Diameter & SEFD\footnote{System Equivalent Flux
Density} \\
                                & (m)      & (Jy) \\
     \hline
    Effelsberg (Germany)  & 100 & 19  \\
    Hartebeesthoek (South Africa) & 26 & 450 \\
    Jodrell Bank Mk2 (United Kingdom) & 25 & 320 \\
    Medicina (Italy) & 32 & 582 \\
    Nanshan (P.R. China) & 25 & 1068 \\
    Noto (Italy) & 32 & 784 \\
    Onsala (Sweden) & 25 & 390 \\
    Sheshan (P.R. China) & 25 & 1130 \\
    Toru\'{n} (Poland) & 32 & 230 \\
    Westerbork (The Netherlands) & 93\footnote{equivalent diameter of the phased
array} & 30 \\
    \hline
    \vspace{-10mm}
    \end{tabular}
    \end{minipage}
\end{table}

\begin{figure*}
  \includegraphics[clip=,bb=30pt 322pt 555pt 560pt,width=150mm, 
angle=0]{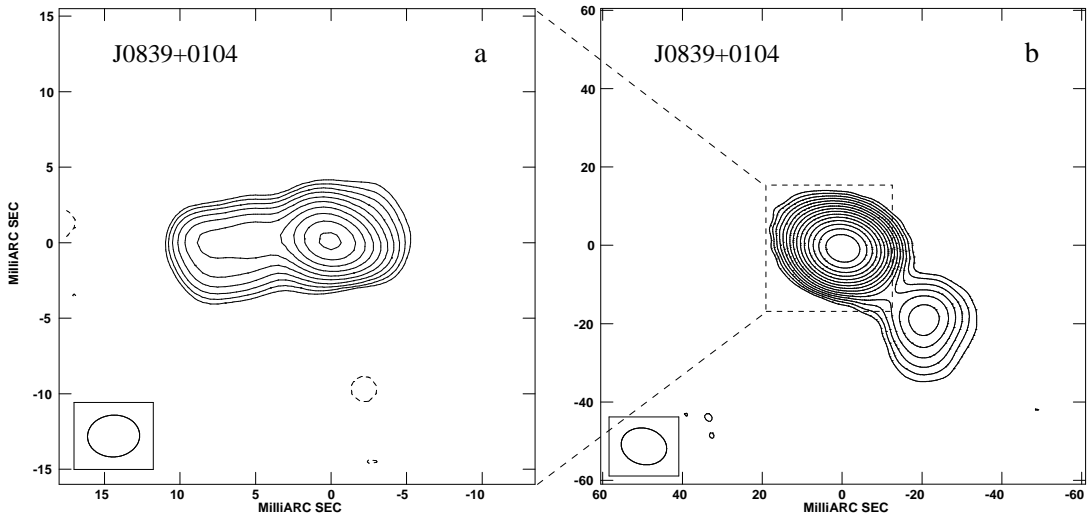}
  \caption{
1.6-GHz VLBI images of the phase-reference calibrator PKS 0837+012 (J0839+0104)
with {\bf (a)} uniform and {\bf (b)} natural weighting. The positive
contour levels increase by a factor of $\sqrt2$. 
{\bf (a)} The first contours are drawn at $-4.5$ and 4.5~mJy/beam. The peak
brightness
is 112~mJy/beam. The restoring beam is 3.5~mas~$\times$~2.8~mas at a position
angle PA=$-87{\degr}$.
{\bf (b)} The first contours are drawn at $-1.3$ and 1.3~mJy/beam. The peak
brightness
is 234~mJy/beam. The restoring beam is 11.6~mas~$\times$~9.2~mas at
PA=$74{\degr}$. 
   }
  \label{calibrator}
\end{figure*}

The NRAO AIPS package \citep[e.g.][]{dia95} was used for initial data
calibration and
imaging. A priori calibration of visibility amplitudes was done using measured
system temperatures for all antennas. The data for the calibrator and
fringe-finder sources (J0839+0104 and J0825+0309, respectively) were
fringe-fitted in AIPS
using 3-min solution intervals. The solutions were interpolated and applied to
the data of SDSS 0836+0054 as well. We then imaged the phase-reference
calibrator source with the conventional self-calibration and CLEAN technique
\citep[e.g.][]{wal95} using the AIPS tasks {\tt CALIB} and {\tt IMAGR}. The
images obtained with two different data weighting schemes are shown in
Fig.~\ref{calibrator}. 

In the uniform weighting scheme, the weights associated with the data points are
inversely proportional to the local density of visibility data, i.e. the longer
projected baselines are generally overweighted. As a result, the angular
resolution becomes relatively high but the image noise level increases.
On the other hand, natural weighting minimizes the image noise by
associating weights proportional to the power of $-2$ of the RMS noise with
each data point. 

\begin{figure}
\centering
  \includegraphics[clip=,bb=40pt 160pt 600pt 663pt,width=75mm, 
angle=0]{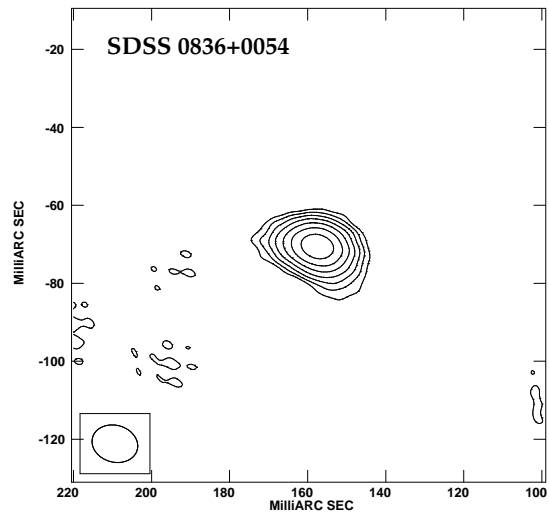}
  \caption{
1.6-GHz VLBI image of SDSS 0836+0054 with natural weighting. 
The positive contour levels increase
by a factor of $\sqrt2$. 
The first contours are drawn at $-75$ and 75~$\mu$Jy/beam. The peak
brightness is 770~$\mu$Jy/beam. The restoring beam is
11.9~mas~$\times$~9.4~mas at PA=$74{\degr}$.
The coordinates are relative to the phase center coincident with the 
optical position.
   }
  \label{target}
\end{figure}

The residual amplitude and phase corrections derived during the hybrid mapping
of the reference source using natural weighting were also applied to the data of
the target quasar.
The naturally weighted image of SDSS 0836+0054 after one cycle of CLEAN 
iterations is shown in Fig.~\ref{target}. The measured
off-source RMS noise in the image is
34~$\mu$Jy/beam, close to the theoretical value.  

\section{Results and discussion}                                                                                                                                             
The quasar SDSS 0836+0054 is clearly detected with VLBI and appears somewhat
resolved at a linear resolution of $\sim70$~pc (Fig.~\ref{target}). 
However, the apparent structural extension may be due to the remaining 
residual phase errors after phase-reference calibration. 

As seen in Fig.~\ref{target}, the position of SDSS 0836+0054 differs from
the optical position reported by \citet{fan01}, but is within the given error 
of 200~mas in both right ascension and declination.
The optical position was used in the correlation of our VLBI data as 
the phase center.
Also within the errors, the coordinates are consistent with the
values taken from the VLA FIRST survey. Our values
$\alpha_{\rm{J2000}}=8^{\rm{h}}36^{\rm{m}}43\fs8606$ and
$\delta_{\rm{J2000}}=0{\degr}54{\arcmin}53\farcs232$ are determined
with respect to the accurately known position of the phase-reference calibrator
J0839+0104. 

However, the differential astrometry is complicated by the fact that the
structure of our reference source, J0839+0104, is rather complex. As seen in
Fig.~\ref{calibrator}b, there is a low surface brightness extended feature to
the SW of the brightest
component. A similar structure in terms of position angle is seen in snapshot
VLBA images taken at higher frequencies (2.3 and 8.4~GHz) in the VLBA
Calibrator Survey \citep{bea02}. With a higher E-W resolution
(uniform weighting, Fig.~\ref{calibrator}a), the
``core'' splits into two distinct components 6.6~mas apart. Supposing that the
jet is one-sided, we suspect that the
true flat-spectrum VLBI core of the quasar is the eastern one which is weaker at
1.6~GHz.
Most likely this is the reference point of which the accurate coordinates are
determined using 8.4-GHz VLBI observations. On the other hand, the brightest
feature in our image
may well be a steep-spectrum jet component which appears stronger than the core
at this lower frequency. It could not be excluded, however, that the brightest
component in Fig.~\ref{calibrator}a coincides with the core. This remains to be
checked with high-resolution VLBI imaging 
of J0839+0104 at frequencies higher than 1.6~GHz.

Since the brightest component of the reference source, which is presumably not
identical with the radio core, is in the phase center of our image, the
relative position determination of the target source may be corrupted by a
systematic error. Therefore we give a conservative error estimate of 8~mas for
the position of SDSS 0836+0054. This reflects the uncertainty in the
identification of the reference point in the radio image of the phase-reference
calibrator source J0839+0104.
  
We searched for possible multiple images of SDSS 0836+0054 that might have been
caused by gravitational lensing. To this end, we investigated a large
($4{\arcsec}\times4{\arcsec}$) field of view centered around the a
priori position of SDSS 0836+0054. To avoid distortion of the field of view due
to frequency-average smearing, we averaged our data over only 8-MHz channels for
imaging. 
We used the correlator output averaging time of 4 s which did not limit 
our field of view. 
No source of compact radio emission has been detected at the brightness level of 
$\sim100$~$\mu$Jy/beam (3 times the image noise), apart from what is seen in
Fig.~\ref{target}. Moreover, the total flux density in the CLEAN component
model corresponding to the naturally weighted image of SDSS 0836+0054
(Fig.~\ref{target}) is 1.1~mJy. This is equal to the flux density in the VLA
FIRST survey. Although the source total flux density recently measured with the
VLA 
at 1.4~GHz is 1.76~mJy (A. Petric et al., in preperation), the difference may 
be attributed to source variability and, to a lesser extent, amplitude
calibration uncertainties in our VLBI data as well. 

Our result suggests that most if not all of the radio emission of SDSS 0836+0054
is confined to a single compact object within an angular extent of $\sim10$~mas.
Thus we can conclude that the quasar is not multiply imaged by gravitational
lensing
at the level of brightness ratio $\loa7$.
Outside of our field of view, there are no 
indications of multiple images neither in the VLA FIRST and NVSS surveys,
nor in the Chandra X-ray image \citep[e.g.][]{sch02}. 
Most recently, \citet{fan03} found that the high resolution ($0\farcs1$) HST
optical image of SDSS 0836+0054 is consistent with an unresolved point source. 
However, it does not exclude
the possibility that its flux is magnified by a factor of up to $\sim2$
\citep{wyi02b}. Note that, based on the size of the ioinized region around 
another SDSS quasar at $z=6.28$, \citet{hai02} place an upper limit of 5 for 
the magnification factor in that case.

Using the two-point radio spectral index of $\alpha=-0.9$ 
(based on VLA total flux density measurements at 1.4 and 5~GHz, A. Petric et
al., 
in preperation; $S\propto\nu^{\alpha}$, where $S$ is the flux
density and $\nu$ is the frequency), the luminosity of SDSS 0836+0054 at 5~GHz 
(rest frame) is $L_5=1.1\times10^{25}$~W~Hz$^{-1}$~sr$^{-1}$. 
Together with the estimated black hole mass of $M_{\rm{bh}}=4.8\times10^{9}
M_{\sun}$
\citep{fan01}, this value is consistent with the $L_{5} \propto
M_{\rm{bh}}^{2.5}$ relation
found for radio quasars at $z\loa1$  
\citep[][and references therein]{jar02}.

\section{Conclusions}                                                            
SDSS 0836+0054, the most distant ($z=5.82$) radio-emitting quasar known to date,
is clearly detected with VLBI observations at 1.6~GHz using the
EVN. A VLBI image with a dynamic range of 10:1 is made (Fig.~\ref{target}).
The radio structure of the quasar at $\sim10$-mas angular resolution is
characterized by a compact component. Using phase-reference
observations to a nearby calibrator, we derived the most accurate astrometric
position of SDSS 0836+0054 available at present. The accuracy of the
coordinates is limited by the complex radio structure of the calibrator quasar
PKS 0837+012, which we imaged for the first time at 1.6~GHz. The extremely
distant quasar SDSS 0836+0054 is unlikely to be multiply imaged by
gravitational lensing. Magnification of its flux due to gravitational lensing
or relativistic beaming cannot be ruled out.

\section*{Acknowledgments}

We thank Chris Carilli for informing us 
about the results of their VLA observations, and the anonymous referee for
useful
suggestions.
The European VLBI Network is a joint facility of European, Chinese and South
African 
radio astronomy institutes funded by their national research councils.
SF, LM and ZP acknowledge partial financial support received from the
Netherlands
Organization for Scientific Research (NWO) and the Hungarian Scientific Research
Fund (OTKA) (grant no. N31721 \& T031723). 
This research was supported by the European Commission's IHP Programme
``Access to Large-scale Facilities'', under contract No.
HPRI-CT-1999-00045. We acknowledge the support of the European Union -
Access to Research Infrastructure action of the Improving Human Potential 
Programme. 
SF and LM acknowledge the hospitality and support of JIVE personnel during the 
scheduling of the observations, and the data correlation and analysis.   
This research has made use of the NASA/IPAC Extragalactic Database (NED) which
is operated by the Jet Propulsion Laboratory, California Institute of
Technology, under contract with the National Aeronautics and Space
Administration.

\bsp

\label{lastpage}

\end{document}